\documentclass[12pt]{article}
\input{amssym.def}
\input{amssym}
\usepackage[dvips]{color}
\usepackage{epsfig}
\usepackage{hyperref}
\normalbaselines
\makeatletter
\@addtoreset{equation}{section}
\makeatother

\def\be{\begin{equation}}
\def\ee{\end{equation}}
\def\ba{\begin{eqnarray}}
\def\ea{\end{eqnarray}}

\newcommand{\rf}[1]{(\ref{#1})}

\def\bra#1{\langle #1|}
\def\ket#1{|#1\rangle}

\begin{document}

\title
{Correlation functions in conformal invariant stochastic processes}

\author{Francisco C. Alcaraz$^1$   and 
Vladimir Rittenberg$^{2}$
\\[5mm] {\small\it
$^1$Instituto de F\'{\i}sica de S\~{a}o Carlos, Universidade de S\~{a}o Paulo, Caixa Postal 369, }\\
{\small\it 13560-590, S\~{a}o Carlos, SP, Brazil}\\
{\small\it$^{2}$Physikalisches Institut, Universit\"at Bonn,
  Nussallee 12, 53115 Bonn, Germany}}
\date{\today}
\maketitle
\footnotetext[1]{\tt alcaraz@if.sc.usp.br}
\footnotetext[2]{\tt vladimir@th.physik.uni-bonn.de}

\begin{abstract}
We consider the problem of correlation functions in the stationary states 
of one-dimensional stochastic models having conformal
invariance. If one considers the space dependence of the correlators, the novel aspect is that although one considers systems with periodic boundary conditions, the observables are described by boundary operators. From our experience with equilibrium problems one would have
expected bulk operators. Boundary operators have correlators having critical exponents being half of those of bulk operators. If one studies the space-time dependence of the two-point function, one has to consider one boundary and one bulk operators.
The Raise and Peel model has conformal invariance as can be shown in the spin 1/2 basis of the Hamiltonian which gives the time evolution of the system. This is an XXZ quantum chain with twisted boundary condition and local interactions. This Hamiltonian is integrable and the spectrum is known
in the finite-size scaling limit. In the stochastic base in which the process is defined, the Hamiltonian is not local anymore. The mapping into an SOS model, helps to define new local operators. As a byproduct some new properties of the SOS model are conjectured. The predictions of conformal invariance are discussed in the new framework and compared with Monte Carlo simulations.
\end{abstract}

\section{ Introduction} \label{sect1}

Many-body local one-dimensional stochastic processes have received a lot 
of attention in both experimental \cite{AAA} and especially theoretical research.
%[BBB]. 
A special class of models are those who have conformal invariance as a 
symmetry. They have a dynamic critical exponent $z = 1$. Conformal 
invariance  gives a lot of information about the expected 
behavior of the system 
in the finite-size scaling limit (large sizes and times) describing the 
approach to the stationary state) \cite{CCC}. Another important prediction of 
conformal invariance is about the expected behavior of the correlation functions at
criticality. One takes the size of the system to infinity first and look at the large 
distance behavior of the correlators next. For equilibrium systems the 
problem is well understood but this is not the case for stochastic 
processes and it is the aim of this paper to clarify it.
 
We will explain why the generalization to non-equilibrium processes of 
correlators is not a trivial matter. When studying one-dimensional 
equilibrium quantum systems with periodic boundary conditions, one takes average quantities in the ground-state
of the system: $\bra{0}F(x_1,t_1)F(x_2,t_2)\ket{0}$ where $F$ is a local 
operator 
(in general a primary field), $\ket{0}=\bra{0}^+$  is the ground-state wave function of
the Hamiltonian which gives the time evolution of the system. For 
simplicity, one can see the Hamiltonian as describing a quantum chain. The 
space-time behavior of the correlator for a periodic system is
 \be\label{eq1.1}
\bra{0} F(x_1,t_1)F(x_2,t_2)\ket{0} = A/R^{2x_b},  
\ee
where $R^2 = (x_2 -x_1)^2 + (t_2 -t_1)^2$ and $x_b$  is 
a critical exponent ($x_b=\Delta +\bar{\Delta}$, where $\Delta$ and 
$\bar{\Delta}$ are  the scaling dimensions). 
 The scaling dimensions are known from the representations of the 
Virasoro algebra which describe the system in the continuum space 
and time in the finite-size scaling limit. The space and time define a 
complex plane or an infinite cylinder of a very large radius. For an open 
system one considers a half-plane respectively a strip.

The situation is very different in the case of stochastic processes. We 
remind the reader that in the continuum time description, the Hamiltonian 
with matrix elements $H_{a,b}$ ($a,b = 1,2,\ldots,N$) is special: 
the non-diagonal 
elements $H_{a,b}$ are non-negative (they are the rates for the transitions 
$b\to a$)
and the diagonal ones are fixed by the non-diagonal ones:
$H_{a,a} = - \sum_{b, b\neq a} H_{a,b}$. A matrix with this property is called an 
intensity matrix and we will call the basis in which $H$ is an intensity 
matrix, a stochastic basis. We stress that a change of basis through a 
similarity transformation changes in general,  the matrix from an intensity matrix to an  
usual matrix. 
In the stochastic basis, the bra vector $\bra{0}^s = \bra{1,1,1,\ldots,1}$ has 
the property 
$\bra{0}^sH = 0$, for any value of $N$.  The time evolution of the system is given by the master 
equation:
\be \label{eq1.2}
\frac{d}{dt} P_a(t) = -\sum_bH_{a,b}P_b(t). 
\ee
$P_a(t)$ gives the probability to find the state $a$ at the time $t$.
For the stationary state, the probabilities $P^0_a$ give a ket-vector 
$\ket{0}  = 
(P^0_1, P^0_2,...,P^0_N)$ which is  an eigenvector: $H\ket{0} = 0$. 
Notice 
that $\bra{0}^s \neq \ket{0}^+$. 

To clarify the picture, it is useful to have the following scenario in 
mind. Assume that one has a non-stochastic basis $\ket{\alpha}$ ($\alpha = 
1,2,\ldots,N)$ in which $H$ is Hermitian and describes an integrable quantum 
chain with local interactions and periodic boundary conditions. We also 
assume that the spectrum of $H$ is known and that in the finite-size 
scaling limit is given by Virasoro representations with a central charge 
$c = 0$ (keep 
in mind that in a stochastic process the ground-state energy is zero for 
any system size). Any correlator of local operators can be computed in the 
standard way and one obtains expressions like \rf{eq1.1}. 
If now we make the change
of basis to the stochastic one, the spectrum stays unchanged, but the 
action of the Hamiltonian in the stochastic basis is not local anymore but 
stays periodic. We will come back to this important point later in the text.
The definition of a correlator in the stationary state is: 
\be \label{eq1.3}
 \bra{0}^s G(x_1,t_1)G(x_2,t_2)\ket{0},    
\ee
where $G(x,t)$ is a diagonal operator in the stochastic basis. 
Since $<0|^s$ is 
a trivial vector, the expression \rf{eq1.3} looks more like a form factor than 
a correlator. 

To simplify the problem, let us look in the stationary state at  space 
correlators only. 
  The way to solve the problem can be found in a seminal 
paper by Jacobsen and Saleur \cite{DDD}. Their interest was to compute 
some properties of the ground-state wave function of the XXZ chain in 
the context of information theory. They considered the complex half-plane 
with $x$ on the horizontal and $y$ on the vertical axis ($y \geq  0$). 
On the $x$-axis one considers two boundary fields \cite{EEE} 
$G(x_1)$ and $G(x_2)$, and 
assumes that one has  Neumann boundary conditions on the $x$-axis (this
corresponds to the bra vector $\bra{0}^s$ in the quantum chain).
 One makes a conformal transformation of the half-plane to
a half-cylinder (a cylinder with a cut perpendicular to the cylinder 
axis). The $x$-axis  is transformed in the circle which cuts the 
cylinder with a coordinate $u$ and the $y$-axis is transformed in the $t$ 
(time) half axis along the cylinder. The time $t$ evolution of the system 
is given by the Hamiltonian in the stochastic basis. One assumes that the 
initial state of the stochastic system is taken at $t = -\infty$ and that the 
system ends up in the stationary state at $t = 0$ (the circle at the cut). 
The boundary operators $G$ become the observables in the stochastic process 
and their correlator can be computed using the conformal field results for 
boundary operators in the half-plane.
Since the correlation functions of 
boundary operators have the critical exponents given by only one Virasoro 
algebra and not two, like for bulk operators, it implies, roughly 
speaking that the critical exponents seen in stationary states have half the 
values of those of equilibrium states.

In Sec.~2 we extend this observation to the non-periodic open system 
and to time correlators. To compute the latter, one needs to consider 
boundary and bulk operators.

We come now to the applications of conformal invariance to stochastic 
models. Unfortunately very few are known and they are all versions of the 
Raise and Peel model \cite{PNR,QQQ} that we discuss in this paper. The model is reviewed 
in Sec.~3. One starts with the periodic Temperley-Lieb algebra 
with the parameter $q = \exp(i\pi/3)$ and write the Hamiltonian as a sum 
of its generators. The algebra has several representations which are 
considered in the text. In the spin 1/2 representation one obtains the XXZ
quantum chain with an anisotropy $\Delta = 1/2$ and twisted boundary
conditions. The Hamiltonian is Hermitian and its  spectrum  
is massless and  given by known representations of the Virasoro algebra with  
a central charge $c = 0$, hence conformal invariant.
  The stochastic basis 
is obtained using other representations of the algebra.
 We use three stochastic base. The first one is the link patterns 
one, the second is the Dyck paths, the third is the particle-vacancy 
basis. They are equivalent. For example, the particle-vacancy
representation is obtained taking the slopes of the Dyck paths. A positive
slope is a particle, a negative slope is a vacancy. In these representations,
the Hamiltonian acts non locally and it is not obvious to find 
observables which can be identified with boundary or bulk operators. 
Although it is 
relatively easy to find observables having the proper space  dependence which can be
compared with the predictions of conformal invariance it might be 
very difficult, if not impossible, to get operators with the proper time dependence. In 
other words, it is much easier to identify local boundary operators and 
much harder to identify bulk operators. To take an example, the current 
density was computed and a conjecture for any lattice size was given
in \cite{FFF}. For a given configuration, the current density gets 
contributions from the action of the generators of the 
Temperley- Lieb algebra on all the sites of the system, 
this makes probably, impossible
to look for a proper time dependence which describes a local behavior. The
reason is the following one. 
In the time interval $\Delta t/L$ which is the time 
scale for the continuum limit (sequential updating in Monte Carlo 
simulations) only one generator hits the configuration and not all of 
them. Another problem is related to the computation of the correlators 
using Monte Carlo simulations in the domain $t/L << 1$ ($L$ is the system 
size). For the accessible values of $L$, the times are so short that one ends 
up with results depending on the initial conditions.

In the case of the Raise and Peel model, one could find local operators 
and their space dependence, looking at the SOS (Coulomb gas) model  picture 
of boundary vertex operators. This is the topic of Sec.~4. Based on the 
work of \cite{EEE}, we could identify one operator (called the NCA for non-crossing
arches) and the contact point operator. This identification was possible
using the generating function for counting the total number of arches
crossing the ends of a segment.  As a by-product of our 
research, we present a conjecture for the generating function giving the
probabilities to have $n$ crossings at one end and $m$ at the other end of the
segment. We also show, using the particle-vacancy stochastic basis 
that the particle density operator is also a "good" boundary operator.

Once the boundary operators have been identified, in Sec.~5 we compare 
the predictions of conformal invariance with the data obtained from 
Monte Carlo simulations. 
We also show that unlike the NCA two-point 
correlation  function, the two-contact point correlation has a 
simple analytic expression but vanishes in the large $L$ limit.
The problems which appear when dealing with 
non-locality are outlined.

In Sec.~6 we present our conclusions.

\section{ Conformal invariance and stochastic processes} \label{sectb}
In this section we present the predictions of conformal field theory for 
the one and two-point correlation functions seen in the stationary state
of a stochastic process. These predictions will be compared  with 
various correlators in Secs.~4 and ~5.

 Consider a one-dimensional system with $N$ states (configurations) 
with probabilities $P_a(t)$ ($a = 1,\ldots,N$). The evolution of the system 
in 
continuum time, is given by  the master equation \rf{eq1.2}.

The Hamiltonian $H$ is an intensity matrix (defined in the Introduction).
One gives the probabilities at the initial time, say $t = -\infty$, and the 
system evolves to a stationary state which is reached at the time $t = 0$. 
One is interested to know the correlation functions in the stationary state.
We assume that the $N$ states can be seen as configurations of a one-dimensional
lattice with $L$ sites and that the spectrum of $H$ was computed and 
found that the finite-size scaling spectrum could be organized in
representations of the Virasoro algebra and that hence one has conformal 
invariance. The central charge of the Virasoro algebra is $c = 0$ since in a 
stochastic process the ground-state energy is equal to zero for any 
number of sites. We would like to know what are the consequences of conformal 
invariance on the correlation functions \rf{eq1.3}. The latter can be obtained, as 
discussed in Secs.~4 and ~5, through computer simulations in which one
discretize the time by taking Monte Carlo steps $\Delta t/L$ or by using 
field theory (taking the lattice space to zero). 

In order to find the consequences of conformal invariance on the 
correlation functions, we take the continuum space and time limits and 
consider separately the cases of periodic and open boundary conditions.
\vspace{0.5cm}

a) {\it Periodic boundary conditions}.

To start with, we consider a half-cylinder with its half-axis being the time 
($t < 0$) coordinate (see Fig.~1a). The cylinder is cut at $t = 0$ which 
corresponds to the time when the system reaches the stationary state. The
horizontal circles on which $H$ is defined, have a perimeter $L$. 
We introduce a
second coordinate $u$ on the circles  ($0 <u < L$). The systems evolves from
the far away bottom of the half-cylinder (the initial state) to its top (the 
stationary state). The observables of the stochastic process in the stationary
state ($t = 0$) used in the calculation of the space dependent 
 two-point functions will 
be denoted by $G(u_1)$ and $G(u_2)$ (they are on the top of the half-cylinder).
If one studies time and space two-point correlations,   one observable 
$G(u_1)$ is on 
the top of the half-cylinder  and the second one $B(u,t)$ on its surface.

%----------------------------------------------------
\begin{figure}
\centering
\includegraphics[angle=0,width=0.5\textwidth] {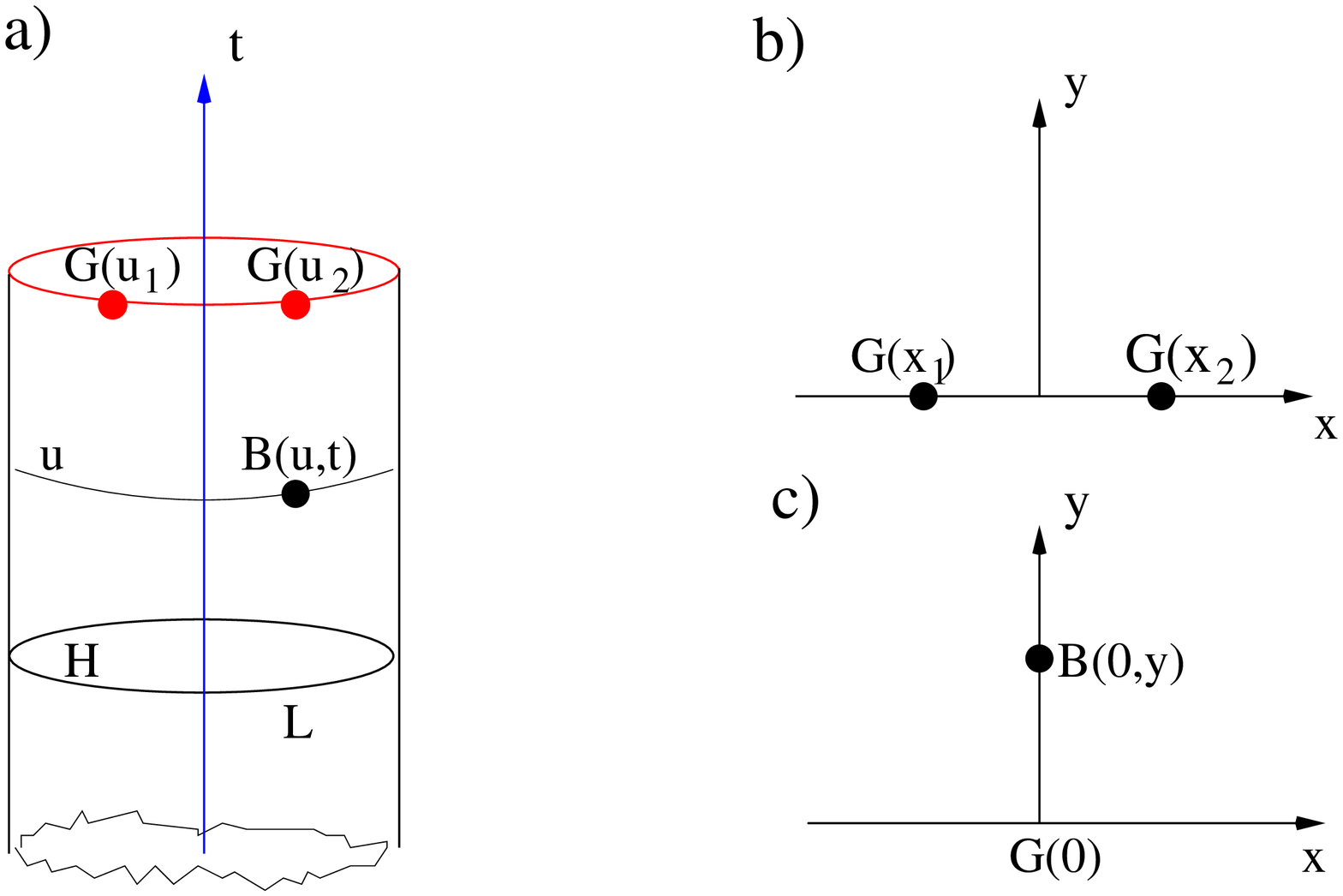}
\caption{
Periodic boundary conditions. a) The observables in the stationary 
state of a stochastic process $G(u_1)$ and $G(u_2)$ are attached to the rim of the
half-cylinder. They are used when computing the space two-point function.
$B(u,t)$ is an observable on the surface of the half-cylinder. It is used when
computing space and time correlators. b) $G(x_1)$ and $G(x_2)$ are boundary fields
in the half-plane. c) $G(0)$ is a boundary field at $x = 0$, $B(0,y)$ is a bulk 
field at $x = 0$ and $y$. 
The Hamiltonian $H$ at time $t$ (negative) acts on the circle of length  $L$  }
\label{fig1a}
\end{figure}
%---------------------------------------------------
%______________________________________________________________

%___________________________________________

In order to get the space correlation functions on the half-cylinder, one 
uses the fact that they are known in the half-plane and that one can make a 
conformal transformation from the half-plane to the half 
cylinder. We denote by $z = x +iy$ ($y\geq 0$) and $w = t +iu$ ($t\leq 0$) 
the complex 
coordinates, and take  Neumann boundary conditions at $y = 0$. This choice 
corresponds to $\bra{0}^s = (1,1,\ldots,1)$ in the stochastic process. The 
correlation function of two boundary fields $G(x_1)$ and $G(x_2)$ (they 
are situated at $y = 0$ and shown in Fig.~1b) is known from conformal field 
theory \cite{MMM}:
\be \label{eqb2}
\bra{0}G(x_1)G(x_2)\ket{0} = |x_2 - x_1|^{-2\Delta}.          
\ee
Here $\Delta$ is the scalling dimension of the boundary field $G$. The Kac 
table provides the following possible values for a $c = 0$ theory:
\be \label{eqb3}
\Delta =\frac{(3r - 2s)^2 - 1}{24},
\ee
where $r$ and $s$ are nonnegative integers ($\Delta = 0, 1/8, 1/3, 5/8, 
1,35/24,\ldots$).
 
The transformation
\be \label{eqb4}
w = \frac{L}{2\pi}\ln \left( \frac{1 + i z}{1 - i z}\right)
\ee
takes the half-plane to the half-cylinder and the two-point correlation 
function on the top of the half-cylinder follows:

\be \nonumber
\bra{0}^s G(u_1)G(u_2)\ket{0} = 
\left(\frac{dx_1}{du_1}\right)^\Delta \left(\frac{dx_2}{du_2}\right)^\Delta \bra{0}G(x_1)G(x_2)\ket{0}.
\ee
Taking $L \to \infty$ with $u_1$ and  $u_2$ fixed, as well  $|u_2 - u_1|$ large, one 
obtains:
\be \label{eqb5}
\bra{0}^s G(u_1)G(u_2)\ket{0} = |u_2 -u_1|^{-2\Delta}.      
\ee
This is an interesting result: although the stochastic process takes place 
in a system with periodic boundary conditions, one doesn't have left and 
right movers. One has only one Virasoro algebra and not two as it is the 
case for periodic boundary conditions in the equilibrium problem.

The one-point function vanishes for large $L$. The large $L$ behavior is 
given by the scaling dimension of the boundary operator. Assuming that on 
the half-line,
\be \label{eqb5-1}
\bra{0}^s G(x)\ket{0} = C, 
\ee
where $C$ is a constant and making the mapping \rf{eqb4} one obtains, for large 
$L$,  the one-point function on the rim of the cylinder:
\be \label{eqb5-2}
G_L(u) \sim~ L^{-\Delta}.      
\ee
  In Sec.~4 and ~5 we are going to make repeated use of Eq.~\rf{eqb5-2}.

If one is interested in time correlation functions 
$\bra{0}^s G(0)B(0,t)\ket{0}$ 
(for simplicity both observables have $u = 0$), one looks again at the 
half-plane (see Fig.~1c) taking $x = 0$ (i. e. $G(x=0)$) and $B(x=0,y)$. Like before,  $G$ is
 a boundary field and $B$ is a bulk field. We have to keep in mind that the 
limit $y \to  0$ is singular (the bulk field $B$ doesn't go smoothly to a 
boundary field). One has \cite{EEE}:
\be \label{eqb6}
\bra{0}G(0)B(0,y)\ket{0} = y^{-x_B -\Delta},              
\ee
where $x_B$ is the conformal dimension of the bulk field $B$. 
We perform again the 
conformal transformation from the half-plane to the half-cylinder and get:
\be \label{eqb7}
\bra{0}^s G(u=0)B(u=0,t)\ket{0} = |t|^{-x_B-\Delta}.         
\ee
In order to derive \rf{eqb7} we have taken $L \to \infty$  with $t$ fixed. 
The
expression \rf{eqb6} generalizes in a simple way when $G$ and $B$ 
have different 
space coordinates, say $u_1$ and $u_2$.

\vspace{0.5cm}
{\it b) Open boundaries}.

\begin{figure}
\centering
\includegraphics[angle=0,width=0.5\textwidth] {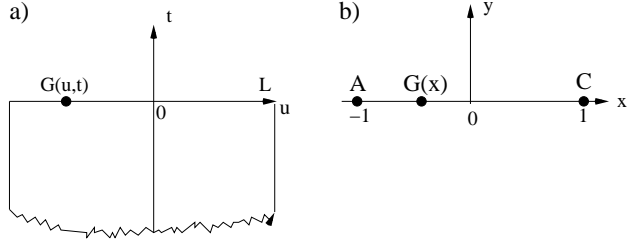}
\caption{
 Open system. a) the strip $w = t + iu$ on which the stochastic process 
takes place. The stationary state is at $t = 0$. The boundary operator  
$G(u)$ is
attached on the top of the strip. b) The half-plane $z = x + iy$ in which 
the conformal field theory is used. $C(-1)$ and $C(1)$ are boundary 
changing fields applied at $x = - 1$ and $x = 1$.}  
\label{figb2}
\end{figure}

If the stochastic process has open boundaries, the problem is slightly 
more complicated. The system is not defined on a half-cylinder but on a 
half-strip of width $L$ (see Fig.~\ref{figb2}). We consider the case of only one 
observable in the stationary state and study the finite-size scaling 
effects on the one-point function $G_L(u)$ ($u = 0,\ldots,L$). The two-point 
correlation function implies the knowledge of model dependent correlations 
in the half-plane.

%----------------------------------------------------
%---------------------------------------------------

In order to get the expression of the one-point function, one consider the 
half-plane with the $y = 0$ axis divided by the points $A$ ($x = - 1$) and 
$C$ ($x 
= 1$). Between $A$ and $C$ ( $- 1 < x < 1$) one takes  Neumann boundary 
conditions. For $x < - 1$ and $x > 1$, we don't need to specify the boundary 
conditions except that they are the same. We attach to the points $A$ and $C$, 
boundary changing fields \cite{EEE} $C(-1)$ and $C(1)$. In order to find $G(x)$, we 
consider the vacuum expectation of the 3-point function given by conformal 
filed theory:
\be \label{eqb8}
\bra{0}\phi_a(x_1)\phi_b(x_2)\phi_c(x_3)\ket{0}> =          
\frac{C_{abc}}{(x_2-x_1)^{\Delta_a+\Delta_b-\Delta_c}(x_2-x_3)^{\Delta_b+\Delta_c-\Delta_a}(x_3-x_1)^{\Delta_a+\Delta_c-\Delta_b}},
 \ee
where $\Delta_a$, $\Delta_b$ and $\Delta_c$ are the scaling  dimensions 
of $\phi_a$, $\phi_b$ and $\phi_c$. Taking $\Delta_a = \Delta_c$, 
$\Delta_b = 
\Delta$  and using \rf{eqb8} one gets:
\be \label{eqb9}
\bra{0}C(-1)G(x)C(1)\ket{0} \sim (1 - x^2)^{-\Delta}.              
\ee
We use the Schwarz-Christoffel mapping which takes the half-plane to the   
strip: $z = - \cosh(\pi w/L)$ ($z = x + iy$, $w = t + iu$) and obtain
\be \label{eqb10}
G_L(u) = \left(\frac{dz}{dw}\right)^{\Delta}\bra{0}C(-1)G(x)C(1)\ket{0} 
       \sim  \left(\frac{L}{\pi} \sin(\pi u/L)\right)^{-\Delta}.
\ee
This result is interesting since there is a theorem \cite{MMM} which says 
that
in equilibrium, the one-point function vanishes unless one has a left-hand 
mover with a scaling dimension equal to the one of the right-hand mover 
therefore one has:
\be \label{eqb11}
G_L(u)_{\mbox{\scriptsize{eq}}} \sim \left(\frac{L}{\pi}\sin(\pi u/L)\right)^{-
2\Delta}.
\ee
Notice the exponent $\Delta$ in \rf{eqb10} as compared to $2\Delta$ in \rf{eqb11}.

  For not symmetric boundary conditions, similar to the equilibrium problem 
\cite{TTT}, Eq.~\rf{eqb10} generalizes as follows:
\be \label{eqb11-1}
G_L(u)\sim F(\cos(\pi u/L)) \left(\frac{L}{\pi} \sin(\pi u/L)\right)^{-\Delta}.
\ee

The time dependence of the one-point function can also be easily 
obtained but for the application we have in mind \rf{eqb10} is enough.

We have to stress that these results were obtained assuming that in the 
stochastic process one uses observables which are local. As we are going
to see in the next sections, to find local observables is not an easy 
task.
 
In Sec.~4 and ~5 we compare the expressions \rf{eqb5}, \rf{eqb5-2}, \rf{eqb10}
 and \rf{eqb11-1} 
with the correlation functions seen in the Raise and Peel model. This 
model is described in the next section.

\section{ The Raise and Peel model} \label{sect2}

There are plenty of papers \cite{PNR} describing this model, written 
after the discovery 
by Razumov and Stroganov \cite{RSS}, that its stationary state has magic 
combinatorial properties. We give here a short survey of the model which 
should allow the reader to follow the results presented in Secs.~4 and ~5.

  One considers a Hamiltonian given by the expression:
\be \label{eqb1}
  H = \sum_{i =1}^M(1 - e_i),
\ee 
where $M = L-1$, for the open system and $M = L$ for the periodic one. $e_i (i=1,\ldots,M)$ 
are 
the generators of the Temperley-Lieb algebra for the open system and of 
the periodic Temperley Lieb algebra for the periodic one. They verify the 
relations
\ba \label{eq2.32}
&&  e_i^2 = (q + 1/q) e_i \\    
&&e_i e_{i\pm 1}e_i = e_i \label{eq2.33}\\
&& e_i e_k = e_k e_i, \quad |i-k| >1. \label{eq2.34}
\ea

In the case of the periodic algebra, which is infinite dimensional,
 another 
relation is needed in order to get a finite dimensional quotient \cite{NNN}. We 
skip it here. The algebras have a representation in terms of Pauli 
matrices and one obtains a Hamiltonian given by an XXZ quantum chain with 
$L$ sites having the $U_q(sl(2))$ symmetry for the open system. 
In this case the  Hamiltonian 
is not Hermitian. In the periodic case, one gets the XXZ chain with 
twisted boundary conditions where  the twist depends on $q$. $H$ is Hermitian in this
case. The spectra of the Hamiltonians are known and for $|q| = 1$, they 
are gapless. Taking $q = \exp(i\pi/3)$ the ground-state eigenvalue is zero for any
value of $L$, and 
the other energy levels are positive and therefore one can look for stochastic base.

\begin{figure}
\centering
\includegraphics[angle=0,width=0.5\textwidth] {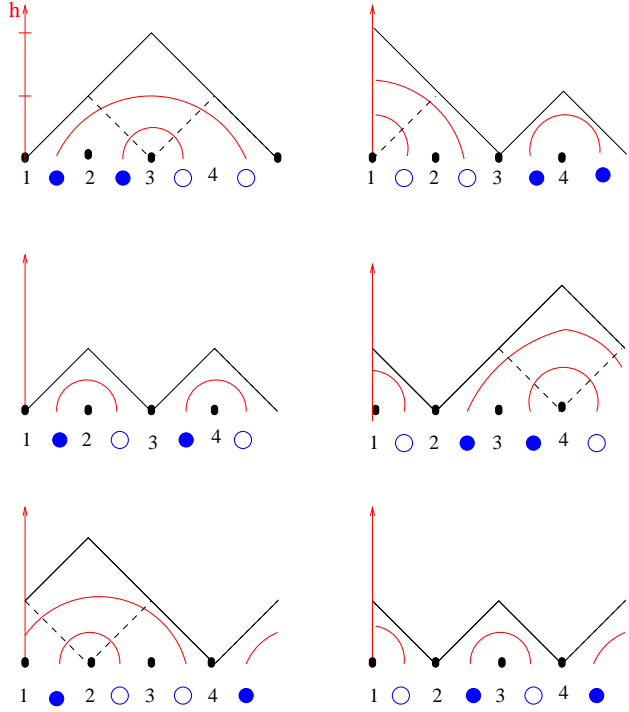}
\caption{
The three stochastic base, Dyck paths, particle-vacancy and link patterns base for 
$L = 4$. There are 
 6 possible profiles.  
In the particle-vacancy representation the particles are  full dots 
and the vacancies empty dots. The  corresponding link patterns  representation 
are also shown.}
\label{fig2}
\end{figure}

  We consider the case $L$ even only. There are three equivalent stochastic 
base, each one useful for different purposes. The simplest one is 
given by Dyck paths. They are defined in terms of heights $\{h_i\}$ obeying the
restricted solid-on-solid rules:
\be \label{eq2.35}
  h_{i+1} - h_i = \pm 1, \;  i = 0,1,\ldots ,L,  
\ee
with $h_0 = h_L = 0$ for the open system and $h_i = h_{L+i}$ for the periodic 
one. In Fig.~\ref{fig2}  we show the six configurations in the periodic 
case $L = 4$. 
There are $L!/\{(L/2 + 1) ([L/2]!)^2\}$  configurations for the open system and 
$L!/[(L/2)!]^2$ for the periodic case.
%----------------------------------------------------
%---------------------------------------------------
%----------------------------------------------------
\begin{figure}
\centering
\includegraphics[angle=0,width=0.5\textwidth] {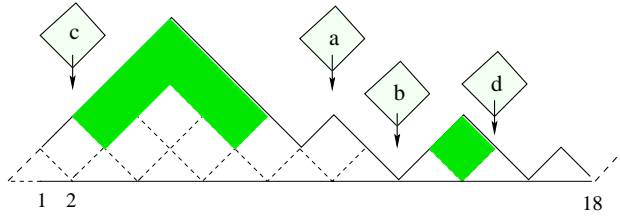}
\caption{
Example of a configuration with four peaks and three contact points in the 
periodic Raise ans Peel model with $L=18$ sites. Depending on the position 
where the tilted tiles reach the interface, several distinct processes 
occur (see the text).}
\label{fig1}
\end{figure}
%---------------------------------------------------

  In the stochastic basis of Dyck paths the time evolution of the 
system can be visualized in the following way. A Dyck path can be seen as 
separating droplets of tilted tiles from a rarefied gas of tilted 
tiles (see Figure \ref{fig1}). The Dyck path changes its shape as a result of 
hits from the tiles of the gas according to the following rules:

In 
a time interval $\Delta t$, with probability $p_i=\Delta t/L$ a given site 
$i$, with height $h_i$ is reached by a tilted tile of the 
gas phase  (see Fig.~\ref{fig1}). Several movements are possible depending 
on the local slop $s_i=(h_{i+1} - h_i)/2$ at the site: 
 
{\it i) $s_i=0$ and $h_i>h_{i-1}$.} 

The tile hits a local peak (case $a$ in Fig.~\ref{fig1}) and it is reflected leaving the profile unchanged.

{\it ii) $s_i=0$ and $h_i<h_{i-1}$.} 

The tile hits a local minimum (case $b$ in Fig.~\ref{fig1}). The tile is absorved ($h_i \to h_i+2$).

{\it iii) $s_i=1$.} 

The tile is reflected after triggering the desorption of a layer of 
tiles from the segment $h_j>h_i=h_{i+b}$, $j=i+1,\ldots,i+b-1$, 
i. e., $h_j \to h_j-2$ for $j=i+1,\ldots,i+b-1$ (case $c$ of Fig.~\ref{fig1}). The desorbed layer contains 
$b-1$ tiles (always an odd number).

{\it iv) $s_i=-1$.}

The tile is reflected after triggering the desorption of a layer of tiles 
from the segment $h_j>h_i=h_{i+b}$, $j=i-b+1,\ldots,i-1$, i. e., $h_j \to h_j-2$ 
 for $j=i-b+1,\ldots,i-1$ (case $d$ of Fig.~\ref{fig1}).

 These evolution rules  take place in  discrete time by choosing, for example,  
$\Delta t=1$ as usual in Monte Carlo simulations or in continuum time by 
taking $\Delta t =1/L\to 0$. 

There is a one-to-one correspondence among  the Dyck path configurations $\{h_i\}$ and the configurations $\{n_i\}$ of particles ($n_i=1$) and vacancies ($n_i=0$) 
attached to the links of the $L$ sites chain, namely,
\be \label{eq2.2}
 n_i = \frac{h_{i+1}-h_{i} +1}{2},\quad (i=1,2,\ldots,L), \quad h_{L+1}=h_1.
\ee
For illustration the configurations for the $L=4$ sites system are also 
shown in Fig.~\ref{fig2}. This correspondence imply that the number of 
particles and vacancies are equal to $L/2$. The time-evolution rules 
of this half-filling chain of particles are directly obtained from the 
corresponding rules for the heights we defined (see \cite{FFF}). In this 
particle language we have a nonlocal asymmetric exclusion model where 
the particles either jump to the next-nearest neighbor position  
(corresponding to  adsorption) at its left, or jump nonlocally to a 
site position on its right (corresponding to  desorption).

We have just described the second stochastic basis, the 
particle-vacancy one. This representation as well as the Dyck path one
correspond to the $S^z = 0$ sector of the XXZ quantum chain.  It can be enlarged
to the full $2^L$ dimensional representation as shown in \cite{FFF}.
A third stochastic basis is the link patterns basis. It is obtained from the Dyck paths basis in the following way. For each height $h_i$ ($i = 1,...,L$) 
one draws $h_i$ noncrossing arches which end up at different sites. 
When all sites $i$ are considered, arches end up at all sites. In Fig.~4 
we illustrate the correspondence between the two base for $L = 4$. 
The dynamics of the system in the link patterns basis follows the one 
in the Dyck path basis.

\section{Connection with the loop gas and the SOS  models}\label{sect3}

\begin{figure}
\centering
\includegraphics[angle=0,width=0.5\textwidth] {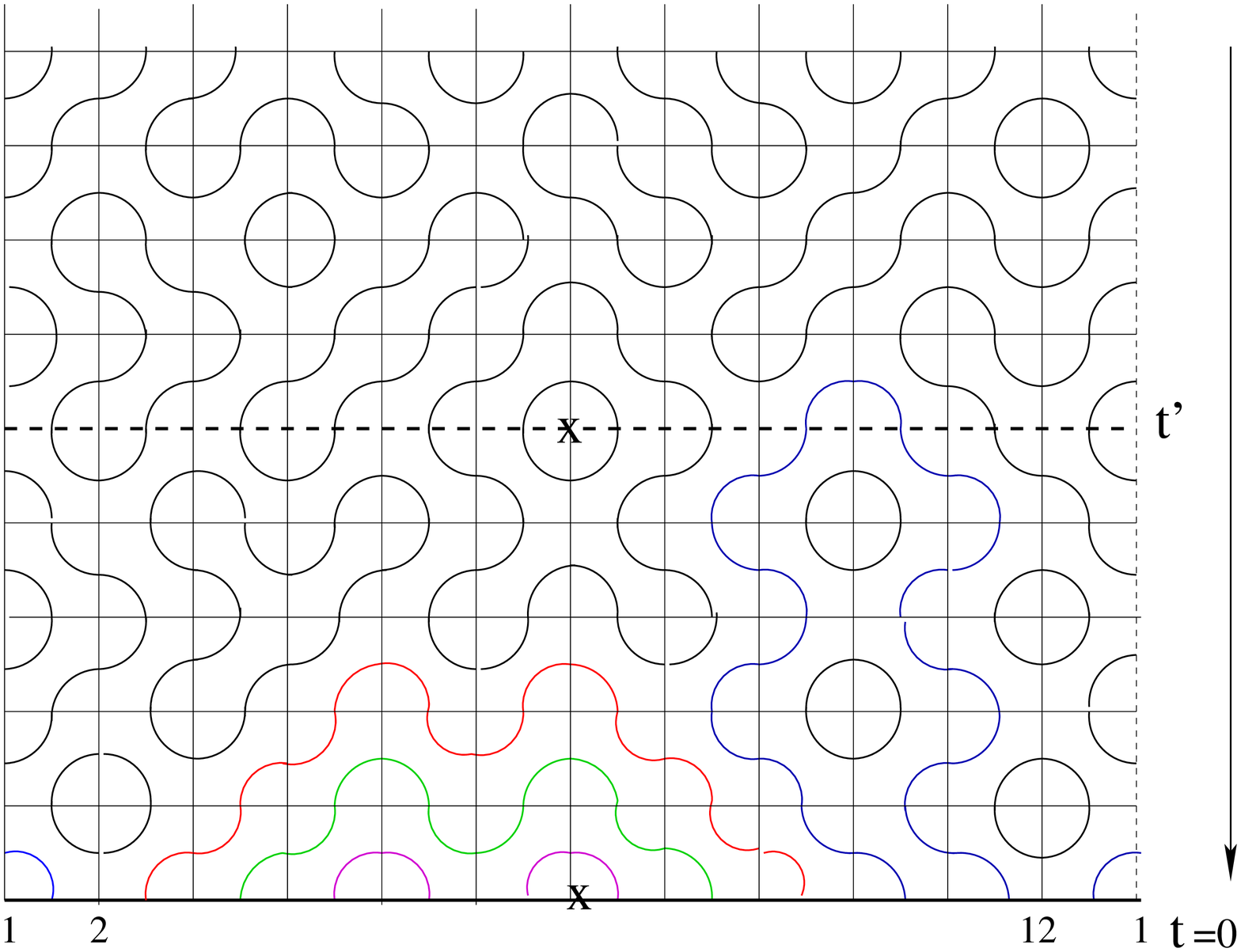}
\caption{
A configuration of self and mutually avoiding fully packed loop model with 
$L=12$ sites in the periodic spacial direction (horizontal). The time direction 
is the vertical direction. The closed arches in the bulk (black loops) as 
well the closed arches (colored arches) that starts and close at the surface (time $t$) have a Boltzmann weight 1.}   
\label{sos1}
\end{figure}

The stationary state of the Raise and Peel model with periodic boundary 
conditions in the link patterns basis can be understood in terms of a $Q = 1$
loop gas model \cite{EEE} on a half-cylinder (see Figs.~\ref{fig1a}  and \ref{sos1}).
 We have in mind
the discussion in Sec.~2 about the applications of conformal invariance.
The rim of the cylinder corresponds to the stationary state. We give 
here a short summary of this connection (see \cite{EEE,DDD} for more details).

One takes a lattice on the half-cylinder and consider only the configurations
of self and mutually avoiding fully packed loops. The configurations of 
the loop model are either closed loops (they don't reach the rim of the
cylinder at $t = 0$) shown in black, or arches which start and end on the rim,
colored in Fig.~\ref{sos1}. All arches have a fugacity equal to 1. The probability to 
have a certain configuration in the stationary state of the stochastic 
model is proportional to the partition function on the half-cylinder 
with the given configuration as a boundary condition. This observation 
is relevant since one can map the loop model into a solid-on-solid (SOS) 
model on the lattice. In the continuum space and time limit, the latter 
model is described by a free bosonic conformal field theory on the rim of the 
half-cylinder \cite{EEE}. The model is parameterized by the parameters $e_0$ equal 
to 1/3 and the coupling $g = 2/3$ in our case. If we denote by $G(u)$ 
the boundary
bosonic field (with Neumann boundary conditions on the boundary) the 
correlator is
\be \label{eq4.1}
<G(u)G(u')> = -3/2\ln|u-u'|^2.
\ee

\begin{figure}
\centering
\includegraphics[angle=0,width=0.5\textwidth] {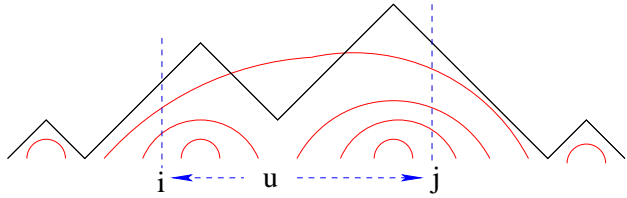}
\caption{
 A configuration with the segment ($i,j$). $h_i = 2$, $h_j = 3$. One 
arch enters the segment at $i$ and leaves it at $j$. }
\label{sos2}
\end{figure}

We define the vertex operators $V_{\pm}(u)= \exp(\pm i(e_1 + e_0/2)G(u))$. 
The use 
 of the new parameter $e_1$ will be clear soon. The correlator of two vertex 
operators is
\be \label{eq4.2}
<V_+(u)V_-(u')> \sim  |u - u'|^{-2\Delta},
\ee
with 
\be \label{eq4.3}
\Delta = (36e_1^2 - 1)/24.
\ee
We can now use Eqs.~\rf{eqb5-1}, \rf{eqb5-2}, \rf{eq4.3}  and get for 
large values of $L$:
\be \label{eq4.4}
<V_{\pm}> = A(e_1)L^{-\Delta}.
\ee 
$A(e_1)$ is a function which has to be determined. Consider now the 
generating function 
\be \label{eq4.5}
M(w) = \sum_{\{h_i\}} P_{\{h_i\}} w^{h_i},
\ee
where $P_{\{h_i\}}$
 is the probability to have a height $h_i$ (Dyck path 
representation) at the site $i$. 
The site independence of $M(w)$ is a consequence of the 
translational invariance  of the stochastic model. This corresponds to $h_i$ arches crossing the 
link $i$ between the sites $i$ and $i + 1$ (see Fig.~\ref{sos2})

\begin{figure}
\centering
\includegraphics[angle=0,width=0.5\textwidth] {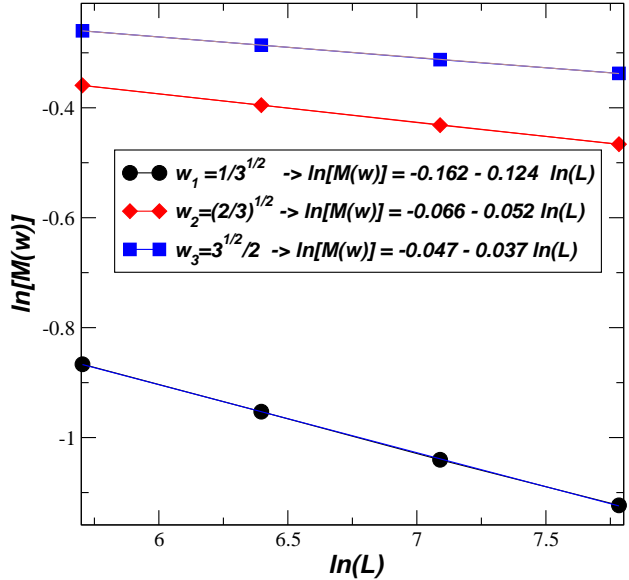}
\caption{
The one-point function \rf{eq4.8} as a function of the lattice size L for the three
values of $w$ given in \rf{eq4.9}. From the fits on can read the values of 
$\Delta$ and $A(w)$.}
\label{fig6}
\end{figure}

It is important to note that following \cite{EEE,DDD} we can get  the  identity
\be \label{eq4.6}
<V_{\pm}> = M(w),
\ee
 where
\be \label{eq4.7}
w = 2\cos(\pi e_1)/\sqrt{3}.
\ee
 In the next section we will discuss the implications of the relation
\be \label{eq4.8}
M(w) = A(w)L^{-\Delta(w)}
\ee
for the Raise and Peel model. Although the relation \rf{eq4.8} might have been 
derived somewhere else, we have checked it using computer simulations.
In Fig.~\ref{fig6} we show the values of the generating function $M(w)$ as a function 
of the lattice size $L$ for 3 values of $w$:
\ba \label{eq4.9}
&&w_1=\frac{1}{\sqrt{3}}, \quad \Delta=\frac{1}{8}=0.125; \quad
w_2= \sqrt{\frac{2}{3}}, \quad \Delta=0.05208333...; \nonumber \\
&& w_3=\frac{\sqrt{3}}{2}, \quad \Delta =0.037720222... \quad . 
\ea
The fits are done using Eq.~\rf{eq4.8} and they are excellent. From the fits we 
obtain
\be \label{eq4.10}
A(1/\sqrt{3}) = 0.850;\quad A(\sqrt{2}/3) = 0.936; \quad A(\sqrt{3}/2) = 0.954 .   
\ee
The value  $A(1)=1$ is a consequence of the normalization of the probability 
function $P_{\{h_i\}}$ in Eq.~\rf{eq4.5}. For completeness we also give the value
\be \label{eq4.11}
A(0) = 2\frac{\Gamma(5/6)}{3\sqrt{\pi}} = 0.424...
\ee        
This value will be derived in the next section.

We consider now the two-point function \rf{eq4.2} and look at the lattice 
realization of it. We take a segment $(i,j)$ on the rim of the 
half-cylinder and consider the generating function:
\be \label{eq4.12}
M'(u;w) = \sum_{h_i,h_j} w^{h_i + h_j}P(h_i,h_j),  
\ee
where $P(h_i,h_j)$ is the probability to have the heights $h_i$ and $h_j$, 
and $u= j -i$. If one looks at the computer simulations one discovers that the 
function $M'(u;w)$ is not independent on $L$ for large lattice sizes. To find 
the proper definition of the generating function, it is better to use the
arches  language. If one considers the segment (see Fig.~\ref{sos2}), 
there are two sorts 
of arches. 
 The first one are those 
 which start at sites smaller than $i$ and end up at sites 
larger than $j$ or starts and ends at the sites between $i$ and $j$. The 
second one are the  arches which start inside and end outside the segment between $i$ and $j$. We call them non-crossing arches (NCA). 
 If we disregard the first kind of 
arches, in the calculation in the new generating function $M(w)$ one finds, 
in agreement with \rf{eq4.2}
\be \label{eq4.13}
M(u;w) = B(w)u^{-2\Delta(w)},
\ee
with
$\Delta$ given by \rf{eq4.3}. This implies $B(1) = 1$, again from the 
normalization of the probability function. The same calculation can be done in
terms of Dyck paths. If we define $h_{\mbox{\scriptsize{min}}}$ as the minimum height in the segment
$(i,j)$
\be \label{eq4.14}
 h_{\mbox{\scriptsize{min}}} =\mbox{Min} \{h_i,h_{i+1},\ldots,h_j\},
\ee
we recover the generating function $M(u;w)$:
\be \label{eq4.15}
M(u;w) = \sum_{h_i,h_j} w^{(h_i - h_{\mbox{\scriptsize{min}}}) + (h_j - h_{\mbox{\scriptsize{min}}})}P(h_i,h_j).
\ee

%---------------------------------------------------
%---------------------------------------------------
%---------------------------------------------------
%---------------------------------------------------

There is an important conclusion coming from this exercise. The 
boundary two-point function of the bosonic field theory has a 
nonlocal lattice realization. Indeed, if we fix the 
segment, the end-points of the segment alone don't fix the paths, one has 
to look at the whole segment and find the value of 
$h_{\mbox{\scriptsize{min}}}$  for each 
configuration. In Sec.~5 we discuss the implications of the connection 
between the generic function $M(u;w)$ and the two-point function \rf{eq4.2}.

The results described up to now can be generalized. Take again the segment 
$(i,j)$ and denote $u = j - i$. Consider the generic function:
\be \label{eq4.16}
M(u;w_1,w_2) = \sum_{\{h_i\}} w_1^{h_i-h_{\mbox{\scriptsize{min}}}}
 w_2^{h_j-h_{\mbox{\scriptsize{min}}}}P(h_i,h_j).
\ee
we conjecture that in the large $L$ limit, we have
\be \label{eq4.17}
 M(u;w_1,w_2) = \frac{C(w_1,w_2)}{u^{\Delta(w_1) + \Delta(w_2)}},    
\ee
where $C(1,1) = 1$. Obviously we recover \rf{eq4.13} if $w_1 = w_2 = w.$ 
 Numerics suggest that $C(w_1,w_2) = A(w_1)A(w_2)$, where $A(w)$ is defined 
by \rf{eq4.8}. 

The conjecture \rf{eq4.17} was checked using Monte Carlo simulations for 
several pairs of values $(w_1,w_2)$. In Figs.~8 and ~9 we show the results 
for the pairs $(1/\sqrt{3},\sqrt{2/3}))$ respectively 
$(\sqrt{2/3},\sqrt{3/4})$.
%----------------------------------------------------
%---------------------------------------------------
%----------------------------------------------------
\begin{figure}
\centering
\includegraphics[angle=0,width=0.5\textwidth] {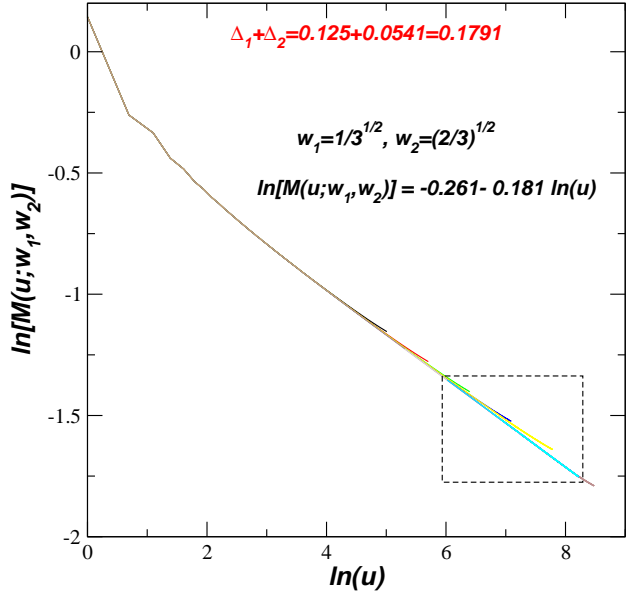}
\caption{
The two-point function   $M(u;w_1,w_2)$ defined in \rf{eq4.17} 
as a function of $u$ for the
values $w_1=1/\sqrt{3}$ and $w_2= \sqrt{2/3}$. The predicted result follows
from  \rf{eq4.9} and is given in the figure.  The lattice sizes are
$L=300,600,1200,2400,4800$ and $9600$. The fitting was done for the
largest lattice and in the
region of the dashed box shown in the figure.}
\label{fig8}
\end{figure}
%--------------------------------------------------
%----------------------------------------------------
\begin{figure}
\centering
\includegraphics[angle=0,width=0.5\textwidth] {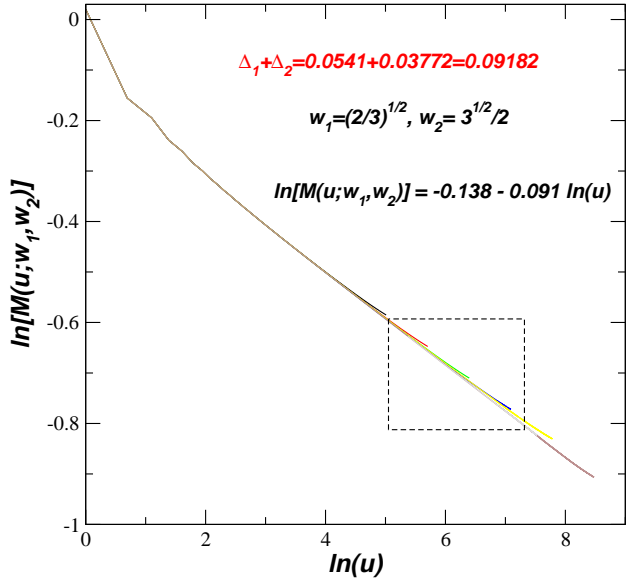}
\caption{
Same as Fig.~\ref{fig8} for the two-point function one  $M(u;w_1,w_2)$ defined in \rf{eq4.17} for the
values $w_1=\sqrt{2/3}$  and $w_2= \sqrt{3}/2$}.
\label{fig9}
\end{figure}
%--------------------------------------------------
We have taken several values of $L$: 300, 600, 1200, 2400, 4800 and 9600. 
One can see that the data are independent on $L$. Moreover, the $u$ 
dependence 
is correctly given by \rf{eq4.17} when using the exponents \rf{eq4.9}. 

We are not aware of any field theoretical derivation of the two-fugacity 
generating function so our results stay as a conjecture. 
%Similarly,  if 
%any results are known from the SOS model for the half-strip we are not 
%aware of them.

\section{ Correlation functions in the Raise and Peel model and conformal invariance}

In Sec.~2 we have given the predictions on conformal field theory for the correlators of stochastic models. These predictions were made for local boundary operators. We have described in Sec.~3 the Raise and Peel model which is conformal invariant but in the stochastic base, the Hamiltonian acts in a nonlocal way therefore the task is to identify local operators in the stochastic base. In Sec.~4 we have shown that the vertex operators of the bosonic boundary field theory are indeed local but that the lattice model from which they can be derived is nonlocal. In this section we are going to clarify the problem and make clearer the connection between the previous sections. We discuss separately the periodic
system and the open one

\vspace{0.5cm}
{\it{a) Periodic boundary conditions.}}

\begin{figure}
\centering
\includegraphics[angle=0,width=0.5\textwidth] {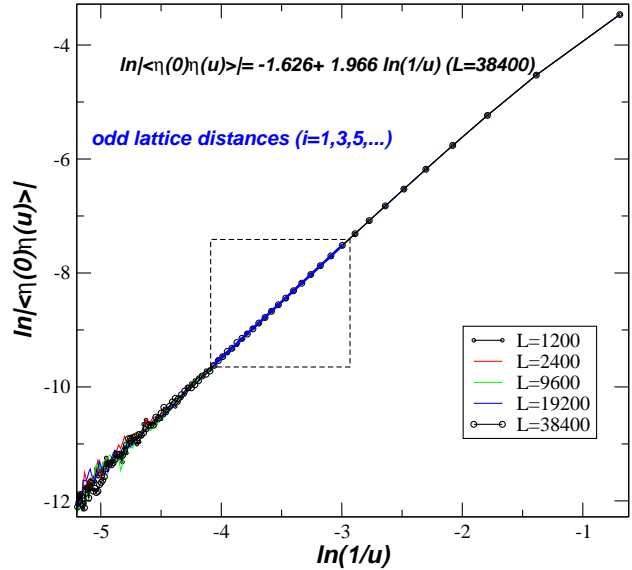}
\caption{
Particle densitity correlator as a function of $u$
Correlation of density of particles for the periodic RPM.
% with density
%of particles $\eta =1/2$ and $u=1$. 
The data are for the lattice sizes  $L=1200,2400,9600,19200$ and $38400$. The plot is in a log-log scale considering
only the odd lattice distances.}
\label{fig10}
\end{figure}

We consider first the particles-vacancy stochastic basis. A natural local variable is the particle density. One expects \rf{eqb5}
 to be valid and this is indeed the case. In Fig.~\ref{fig10} we show the density-density correlator $<\eta(0)\eta(u)>$ obtained from Monte Carlo simulations for different lattice sizes ($\eta(u)$ is 
the density of particles minus 1/2). There is no lattice dependence and one obtains an exponent $\Delta = 0.983$ which within errors is equal to 1, one of the 
exponent  given by \rf{eqb3}. This exponent is the one one expects on dimensional grounds and will also be obtained in the open system.
%----------------------------------------------------
%--------------------------------------------------

\begin{figure}
\centering
\includegraphics[angle=0,width=0.5\textwidth] {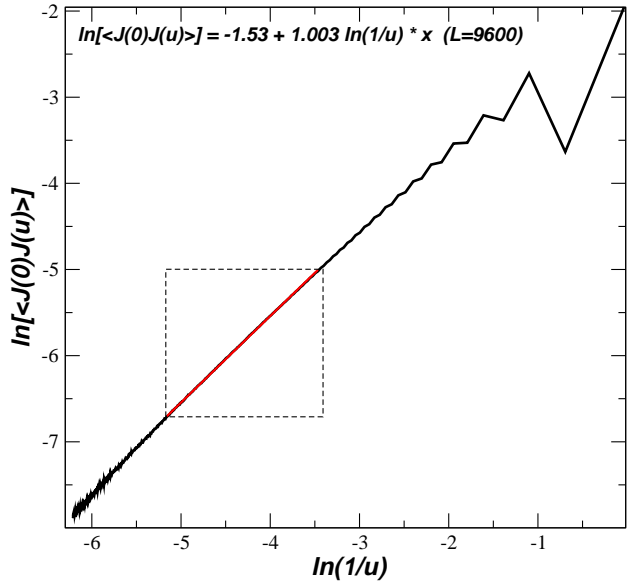}
\caption{
Current density correlator  as a function of $u$ for the periodic RPM.
% with density
% of particles $\eta=1/2$. 
The data are for $L=9600$}
\label{fig11}
\end{figure}

Another local observable is the current density in the stationary state. It was shown in \cite{APP} that its expression for large $L$ is:
\be \label{eq5.1}
J(L) = \frac{3}{4L}.
\ee
One would expect therefore on the two-current correlator $<J(0)J(u)>$ an exponent $2\Delta = 2$. Monte Carlo simulations (see Fig.~\ref{fig11}) give an exponent equal to 1.028 which is close to 1 and not to 2. We have no explanation for this observation.

%----------------------------------------------------
%--------------------------------------------------

There is no obvious connection between the bosonic boundary field theory and the density of particles as well  the current density observables.

We consider now the Dyck paths stochastic basis. A natural local observable is the density $R(L)$ of contact points ($h_i = 0$) (see Fig.~4). There is an exact result for this quantity. It is based on a
conjecture by de Gier \cite{JDG} for the number of clusters for any $L$. A cluster is a segment between two consecutive contact points. This conjecture was verified by Monte Carlo simulations. The expression of the density for large $L$ is;
\be \label{eq5.2}
R(L) =\frac{2}{3}\sqrt{\pi}\Gamma(5/6)/L^{1/3}.
\ee
This gives an exponent $\Delta = 1/3$, again given by \rf{eqb3}. The same result is obtained using \rf{eq4.3} and \rf{eq4.7} with $w = 0$.

The definition of the two-contact-point correlation has to be considered with care. If one uses the SOS prescription discussed in Sec.~4, one has to use 
Eq.~i\rf{eq4.2} for $w = 0$ and get from \rf{eq4.3} and \rf{eq4.7}   $\Delta = 1/3$. 
We have checked this prediction  using Monte Carlo simulations. This implies that one has to take into account not only  the configurations in which $h_i = h_j = 0$ ($j - i = u$) but also all the configurations in which 
$h_i = h_j = h_{\mbox{\scriptsize{min}}}$. Those are the NCA 
configurations.  
It turns out that if we take into account only the configurations with contact points ($h_i = h_j = 0$) the two-point function vanishes in the large $L$ limit but it does so in a neat, unexplained. way:
\be \label{eq5.3}
M'_L(u,0)  \sim (Lu)^{-1/3}.
\ee
The physical interpretation of \rf{eq5.3} is the following one. 
The conditional probability to have a contact point at the distance $u$
 from a given contact point (that happens with probability  $L^{-1/3}$) decreases 
like $u^{-1/3}$.

Before closing this section we give some results which can be obtained using the generating functions $M(w)$ (Eq.~\rf{eq4.8}), $M(u;w)$ (Eq.~\rf{eq4.13}) and 
$M(u;w_1,w_2)$ (Eq.~\rf{eq4.17}) defined in Sec.~4. One obtains in this way 
%parei
properties of the Dyck paths configurations in the large $L$ limit.

From the one-point vertex $M(w)$ (4.8) one obtains
\be \label{eq5.4}
<h^n> = \left. \left(w\frac{d}{dw}\right)^n M(w)\right|_{w=1},
\ee
\be \label{eq5.5}
P(h=n) = <\delta_{h,n}> = \frac{1}{n!} \left. \frac{d^n}{dw^n} M(z)\right|_{w=0}.
\ee
It is easy to derive the leading behavior:
\be \label{eq5.6}
<h> = \frac{\sqrt{3}}{2\pi} \ln{L},
\ee
\be \label{eq5.7}
\sigma_h^2 = <h^2> - <h>^2 = \frac{2\sqrt{3} \pi -9}{\pi^2} \ln{L},
\ee
and
\ba \label{eq5.8}
&&P(h=0) = A(0)L^{-1/3}, \quad 
P(h=1) = A(0)\frac{3\sqrt{3}}{4\pi} \frac{ \ln{L}}
{L^{1/3}}, \nonumber \\
&& P(h=2)= A(0)\left(\frac{27}{32\pi^2}\frac{(\ln L)^2}{L^{1/3}}
-\frac{9}{8\pi^2} \frac{\ln{L}}{L^{1/3}}\right),
\ea
with $A(0)$ given by \rf{eq4.11}. 

From $M(u;w_1,w_2)$ (Eq.~\rf{eq4.16}) we get the leading behavior:
\be \label{eq5.9}
<h_i - h_{\mbox{\scriptsize{min}}}> = \left. \frac{\partial}
{\partial w_1} M(u;w_1,w_2)\right|_{w_1=w_2=1} = \frac{\sqrt{3}}{2 \pi} \ln{u}.
\ee

Notice that in the large $u$ limit $<h -h_{\mbox{\scriptsize{min}}}>$ coincides with (5.6). This implies that the average number of arches crossing a link is equal to the average number of arches entering the segment of length $u$ from one side 
or the other. The double of this quantity is called valence bond bipartite entanglement entropy (see  e.g. in \cite{JLS,DDD}).

An interesting quantity is the average probability to have a height
$(h_i - h_{\mbox{\scriptsize{min}}}) = m$ at the site $i$ and a height 
$(h_j - h_{\mbox{\scriptsize{min}}}) = n$ at the site $j$.
Using \rf{eq4.16} and \rf{eq4.17} one finds the leading behavior:
\be \label{eq5.9b}
P_{m,n} (u) \sim |\ln(u)^{m+n}|u^{-2/3}.
\ee
\vspace{0.5cm}
{\it b) Open boundary conditions.}

We consider first symmetric boundary conditions and a stochastic basis with particles and vacancies in an open segment of length $L$ ($L$ even). Let $u$ 
be the distance from one of the boundaries and $\rho$ the density of 
particles from which one has subtracted 1/2 (the average value). Using Monte Carlo simulations \cite{APR} we have found
\be \label{eq5.10}
\rho = C \frac{\pi}{L\sin(\pi u/L)},
\ee
where $C$ is a constant. Using \rf{eqb10} we find $\Delta = 1$ in agreement 
with the result obtained in the periodic boundary case.

We next consider the density of contact points $R_L(u)$. 
An exact result is known in this case \cite{PNR,APR}:
\be \label{eq5.11}
R_L(u) = C_1 \left(\frac{\pi}{L\sin(\pi u/L)}\right)^{1/3},
\ee
where $C_1 = - \sqrt{3}\Gamma(-1/6)/( 6(\pi)^{5/6}) = 0.7531...$ . Comparing with \rf{eqb10} one finds $\Delta = 1/3$ in agreement with what was obtained in the periodic case.

The case of nonsymmetric boundaries was studied in Ref.~\cite{APR}. The one-boundary Temperley Lieb algebra was used adding a boundary generator to the Hamiltonian \rf{eqb1} and the function $F(\cos(\pi u/L))$ in Eq.~\rf{eqb11-1} was identified for both the densities of particles and of contact points.
%parei :1

\section{Conclusions}

We think that in this paper we have clarified the consequences of conformal invariance in stochastic processes. The main ingredient is the realization that if one looks at space dependent correlators one deals not with bulk operators 
but with Neumann boundary operators.  Since there are no left and right movers in this case, the critical exponents are half of what  one would expect for a periodic system. Space and time correlators imply that one has to consider also bulk operators in the vicinity of the boundary. One conclusion is that one shouldn't expect the
standard relativistic space-time dependence $x^2 + t^2$ of the correlators.

This picture was verified in the Raise and Peel model which has a simple, albeit nonlocal dynamics which allows Monte Carlo simulations on large lattices. We have checked in this way the predictions of conformal invariance
when analytical results were not available.

Because of the nonlocal dynamics, the identification of local operators in the stochastic base for which conformal field theory makes predictions, is not an obvious task. This identification is dependent on which basis was chosen.

In the particle-vacancy basis, we have checked that the particle density is a bona fide local operator with scaling dimension $\Delta = 1$. The current density operator on the other hand has a correlator with an unexplained behavior. In the Dyck paths basis, one can use the mapping of the model onto the SOS model which in continuum space and time
is described by a  bosonic free field theory. Using boundary vertex operators with a known behavior, one can  go backwards and identify the local operators in the lattice model.  

Looking at Dyck paths, the density of contact (zero height) points look as a natural candidate for a local operator. Examining the behavior of the one-point function, it looks that this is the case. One finds a scaling dimension $\Delta = 1/3$. Looking however at the two-contact point correlator, we discover that it vanishes at large values of $L$. The $L$ and $u$ 
(the distance between the two contact points) dependence is simple, it involves again 
the exponent 1/3, but the expression of the correlator remains a puzzle.
The way to solve the problem comes from the bosonic field theory. 
If one takes the segment between the two points where the correlator is computed, one sees that for each configuration one has a minimum height $h_{\mbox{\scriptsize{min}}}$. If one subtracts from each height $h_{\mbox{\scriptsize{min}}}$ 
one gets new contact points for which $h - h_{\mbox{\scriptsize{min}}} = 0$. 
If one takes into account now all the configurations having contact points including also of those of the new kind at the end of the segment one finds a non-vanishing correlator with the proper critical exponent.
This is the NCA (non-crossing arches correlation). 

Let us observe that from all the possible scaling dimensions (2.3) available for a $c = 0$ theory, we found only two of them: 1/3 and 1.

As a byproduct of our research, we made the conjecture (4.12) 
on the average
probability to have two different subtracted ($h_{\mbox{\scriptsize{min}}}$ is taken away) heights at the end of a given segment. This conjecture was checked in Monte Carlo simulations.

As the reader might have observed, we have not touched on the space and time correlators. To our knowledge, exact results are not known and computer simulations on very large lattices have large finite-sizes corrections depending on the initial conditions. We plan to look in the future at this problem.

Before closing the paper, we would like to observe that with the exception of several versions of the Raise and Peel model, no other models with conformal invariance and a simple dynamics are known. This is an invitation for more research in this field

\section{Acknowledgments}
VR would like to thank David Mukamel for his kind invitation to the 
Department of Physics of Complex Systems at the Weizmann Institute where 
part of this work was done.
 and FCA to FAPESP and CNPq (Brazilian agencies) for financial support.

\end{document}